\pdfoutput=1

\documentclass[conference]{IEEEtran}

\usepackage{amsmath}
\usepackage[detect-all]{siunitx}
\usepackage[inline]{enumitem}
\usepackage[capitalize,noabbrev]{cleveref}
\usepackage{graphicx}
\graphicspath{{images/}}
\usepackage{booktabs}
\usepackage{dirtytalk}
\usepackage[most]{tcolorbox}
\usepackage{romannum}
\usepackage{url}

\AtBeginDocument{\pagenumbering{arabic}}

\makeatletter
\def\findingcounter{\thetcb@cnt@finding}
\makeatother

\newtcolorbox[auto counter]{finding}{
	enhanced,
	breakable,
	colback=white!5!white,
	colframe=gray!75!black,
	title=\textbf{Finding~\thetcbcounter}
}

\newcommand{\bsd}{\textsc{BSD}}
\newcommand{\dbsd}{{D}ragon{F}ly{BSD}}
\newcommand{\fbsd}{{F}ree{BSD}}
\newcommand{\nbsd}{{N}et{BSD}}
\newcommand{\obsd}{{O}pen{BSD}}

\newcommand{\cc}{code churn}
\newcommand{\cor}{code review}
\newcommand{\krn}{kernel}
\newcommand{\loc}{\textsc{LOC}}
\newcommand{\nonkrn}{non-kernel}
\newcommand{\os}{operating system}
\newcommand{\oss}{open-source software}
\newcommand{\phab}{{P}habricator}
\newcommand{\tta}{time-to-accept}
\newcommand{\ttfr}{time-to-first-response}
\newcommand{\ttm}{time-to-merge}

\newcommand{\initialcommitcount}{\num{797879}}
\newcommand{\finalcommitcount}{\num{796821}}

\newcommand{\initialfreebsdcrcount}{\num{32884}}
\newcommand{\finalfreebsdcrcount}{\num{14875}}

\title{Is Kernel Code Different From Non-Kernel Code? A Case Study of \textsc{BSD} Family Operating Systems}

\begin{document}

\author{
    \IEEEauthorblockN{Gunnar Kudrjavets\IEEEauthorrefmark{1},
                      Jeff Thomas\IEEEauthorrefmark{2},
                      Nachiappan Nagappan\IEEEauthorrefmark{2},
                      Ayushi Rastogi\IEEEauthorrefmark{1}}

    \IEEEauthorblockA{\IEEEauthorrefmark{1}\textit{Bernoulli Institute}\\
                      \textit{University of Groningen}\\
                      9712 CP Groningen, Netherlands\\
                      Email: g.kudrjavets@rug.nl, a.rastogi@rug.nl}

    \IEEEauthorblockA{\IEEEauthorrefmark{2}\textit{Meta Platforms, Inc.}\\
                      1 Hacker Way\\
                      Menlo Park, CA 94025, USA\\
                      Email: jeffdthomas@fb.com, nnachi@fb.com}
}

\maketitle

\begin{abstract}
Code churn and code velocity describe the evolution of a code base.
Current research quantifies and studies code churn and velocity at a high level of abstraction, often at the overall project level or even at the level of an entire company.
We argue that such an approach ignores noticeable differences among the subsystems of large projects.
We conducted an exploratory study on four \bsd\ family \os s: \dbsd, \fbsd, \nbsd, and \obsd.
We mine \initialcommitcount\ commits to characterize code churn in terms of the annual growth rate, commit types, change type ratio, and size taxonomy of commits for different subsystems (\krn, \nonkrn, and mixed).
We also investigate differences among various \cor\ periods, i.e., \ttfr, \tta, and \ttm, as indicators of code velocity.

Our study provides empirical evidence that quantifiable evolutionary code characteristics at a global system scope fail to take into account significant individual differences that exist at a subsystem level.
We found that while there exist similarities in the code base growth rate and distribution of commit types (neutral, additive, and subtractive) across BSD subsystems,
\begin{enumerate*}[label=(\alph*),before=\unskip{ }, itemjoin={{, }}, itemjoin*={{, and }}]
    \item most commits contain \krn\ or \nonkrn\ code exclusively
    \item \krn\ commits are larger than \nonkrn\ commits
    \item \cor s for \krn\ code take longer than \nonkrn\ code.
\end{enumerate*}
\end{abstract}

\begin{IEEEkeywords}
BSD, code churn, code velocity, kernel code, non-kernel code
\end{IEEEkeywords}

\section{Introduction}

A widely accepted definition for \emph{software evolution} is a continual change from a lesser or worse state to a higher or better state~\cite{arthur_1988,tripathy_2014}.
The essence of this change is a \say{process by which programs are modified and adapted to their changing environment}~\cite{herraiz_2013}.
At the lowest implementation level, evolution of software depends on modifications to lines of code (\loc)~\cite{spinellis_software_2021}.
We use \emph{code churn}~\cite{munson_1983} to quantify software evolution~\cite{hall_software_2000}.

Engineers identify code changes differently depending on historical conventions, a specific project, or a code collaboration tool.
Developers can refer to proposed code changes as diffs, patches, or pull requests.
In this paper, the unit of analysis for code churn is a commit.
Developers use {M}odern {C}ode {R}eview to validate the correctness of proposed code changes~\cite{sadowski_modern_2018}.
As a result, the frequency and speed of software evolution depend on how fast the proposed code changes are reviewed and merged to the target branch.

Existing research about the size-related characteristics of code churn (e.g., size of commits)~\cite{rigby_convergent_2013,sadowski_modern_2018,
rossi_continuous_2016,rigby_convergent_2013,
rigby_2011,gousios_exploratory_2014}
and code reviews~\cite{macleod_2018,bird_2015,izquierdo-cortazar_2017,jiang_2012,zhu_2016,tan_2019} focuses either at the level of an entire project or a company.
For example, the role of patch review in software evolution was investigated in the context of an entire Mozilla Firefox project~\cite{nurolahzade_2009}.

Given the large number of engineers working on more significant software
projects, our exploratory case study proposes that \emph{there exist many subsystems with potentially different subcultures in the context of one large system}.
A subsystem can be one of the many products a company such as Google or Microsoft is developing or even a different architectural layer of the Linux \krn.
Code churn patterns, commit sizes, and code velocity may differ between those subsystems.
\emph{Viewing the entire system as a monolithic entity may lead to incorrect conclusions about the characteristics of a system}.
For example, statements like \say{the median \cor\ size for X is \textit{N}} is overly simplistic, omitting the potentially significant differences between either various products developed by the same company or the components of a more extensive system.

We set forward to find out if there is empirical evidence to show that code churn and code velocity  differ between various subsystems.
We investigate various characteristics of code churn, such as commit size, commit type, and the ratio of different types of code changes.
To quantify the speed at which a system evolves, we measure the various \cor\ intervals: \ttfr, \tta, and \ttm.

The unit of analysis in this paper is a specific type of large software system---\emph{\os s}.
We divide the \os\ code into two categories: \emph{\krn}\ and \emph{\nonkrn}\ code (see~\Cref{subsec:terminology} for terminology and rationale).
This categorization is just one of the possible ways to split the code.
There are many other options to classify various \os\ components.
For example, we can divide an \os\ into subsystems such as a file system, \textsc{I/O} manager, memory manager, or user mode utilities.

We did not find any existing studies comparing code churn and \cor\ velocity between \krn\ and \nonkrn\ code in the context of the same \os.
Therefore, we conduct an exploratory study on a subset of \bsd\ family \os s (\dbsd, \fbsd, \nbsd, and \obsd) to understand if there is any validity to our intuition.
The \emph{need to analyze a complete \os} that contains
everything from various command-line utilities to \krn\ and
windowing systems dictates the choice of various \bsd\ versions.
For example, hundreds of distributions include the Linux kernel.
Because of various distributions, understanding what exactly constitutes \nonkrn\ code for Linux is not clear and depends highly on a distribution.

We pose the following research questions:

\begin{enumerate}
	\item \textbf{RQ1}: are code churn characteristics different between \krn\ and \nonkrn\ code?
	\item \textbf{RQ2}: is there a difference in the code velocity between \krn\ and \nonkrn\ code?
\end{enumerate}

We collect information by mining GitHub mirrors of all the \os s that we study.
To investigate code velocity, we mine \phab~\cite{phacility} instance used for \fbsd\ \cor s.
\phab\ is a modern code collaboration tool used by popular projects such as Mozilla and \textsc{LLVM}.
We report descriptive statistics for all the research questions.
We find that
\begin{enumerate*}[label=(\alph*),before=\unskip{ }, itemjoin={{, }}, itemjoin*={{, and }}]
    \item engineers mainly make changes that are restricted to a specific abstraction level, i.e., they change either \krn\ or \nonkrn\ code and rarely mix both in the same commit
    \item \krn\ commits are larger than \nonkrn\ commits
    \item median annual growth rate and  distribution for neutral, additive, and subtractive commits for \krn\ and \nonkrn\ code are similar
    \item \cor s in \fbsd\ \krn\ code take longer than for \nonkrn\ code.
\end{enumerate*}

\section{Background and Related Work}

\subsection{Related work}

\subsubsection{Code churn size}
\label{subsubsec:code_churn_size}

From industry studies, we know that median change size at \textsc{AMD} is \num{44} lines~\cite{rigby_convergent_2013},
the median number of lines modified at Google is \num{24}~\cite{sadowski_modern_2018},
at Facebook each deployed software update involved, on average, \num{92} lines of code~\cite{rossi_continuous_2016},
and for Lucent, the number of non-comment lines changed is \num{263} lines~\cite{rigby_convergent_2013,porter_1998}.
We observe an \emph{order of magnitude in the code change between companies}.

Including data about \oss\ adds even more variety to our data set.
In \oss\ projects, the median change size ranges from \num{11} to \num{32} changed lines~\cite[p.~37]{rigby_2011}.
For Android, the median change size is \num{44} lines,
for Apache \num{25} lines,
for Linux \num{32} lines, and
Chrome \num{78} lines~\cite{rigby_convergent_2013}.
The median size of pull requests in GitHub is \num{20} total lines changed~\cite{gousios_exploratory_2014}.

The critical observation from available data is that \emph{the size of changes varies significantly even between various \oss\ projects}.
We did not find any existing studies differentiating the size of commits between various layers of abstraction in the same \os.

\subsubsection{Code velocity}

Existing research about code churn and code velocity in \os s focuses on \oss\ such as \fbsd\ or Linux.
Industry researchers generally publish data within the scope of the company.
We are not aware of any \emph{externally published} research related to commercial \os s, such as closed-source derivatives of Darwin (foundation for Apple's \os s such as iOS, macOS, and watchOS) or Microsoft Windows.

Based on current findings, we know that the \cor s at Google, Microsoft, and \oss\ projects take approximately \num{24} hours~\cite{rigby_2013}.
In commercial software development (e.g., Google), \num{24} hours is the general expectation for the \cor\ turnaround time~\cite[p.~176]{winters_2020}.
An immediate observation we make is that this number is an order of magnitude smaller than the time it takes to review the Linux \krn\ code.

Code review practices for the Linux kernel have been studied~\cite{erdamar_measuring_2021}.
However, that research mainly focuses on \cor\ activity and its participants, not specifically on the code churn and velocity.
A study on the Linux kernel finds that \say{33\% of the patches makes it into a Linux release, and that most of them need 3 to 6 months for this}~\cite{jiang_2012}.
Another study analyzing \num{139664} accepted patches in the Linux kernel states that \emph{review time} (from patch being published to its acceptance) is on median \num{20} days~\cite{tan_2019}.
An existing study on \fbsd\ results in findings differing by order of magnitude~\cite{zhu_2016} albeit it is another \textsc{UNIX}-like \os.
The data source is approximately \num{25000} \cor\ submissions from \fbsd\ mailing lists.
The study defines \emph{resolve time} as \say{the time spent from submission to the final issue or pull request status operation (stopped at \emph{committed}, \emph{resolved}, \emph{merged}, \emph{closed}) of a contribution.}
The median resolve time for \fbsd\ is \num{23} hours.

We observe from these data points at least an order-of-magnitude difference in code velocity even in the context of Linux.
The \krn\ code (Linux), company-wide statistics (Google, Microsoft), and overall \os\ scope (\fbsd) have wide variance.
Existing studies show the difference in software evolution for code churn characteristics and code velocity.
We argue that these differences exist even within the same system.

\subsection{Classification, history, and scope of \bsd}

The history of \textsc{UNIX}~\cite{salus_quarter_1994} and \bsd\ family \os s is extensively documented~\cite[p.~3--14]{mckusick_design_2015}.
All the \os s we investigate have a common set of ancestors:
\begin{enumerate*}[label=(\alph*),before=\unskip{ }, itemjoin={{, }}, itemjoin*={{, and }}]
    \item \nbsd\ has its roots in both {386\bsd}\ and {4.4\bsd\ Lite}
    \item \obsd\ is based on \nbsd\ source tree
    \item {\fbsd\ 1.0} is based on {386\bsd\ 0.1}
    \item initial commit for \dbsd\ is based on {\fbsd\ 4.8}.
\end{enumerate*}

Different flavors of \bsd\ have a specific focus.
For example, \obsd\ has a \say{fanatical attention to security, correctness, usability, and freedom} and \say{strives to be the ultimate secure operating system}~\cite[p.~4]{lucas_absolute_openbsd_2003}.
At the same time, \say{NetBSD's main purpose is to provide an operating system that can be ported to any hardware platform}~\cite[p.~\romannum{33}]{lucas_absolute_freebsd_2002}.

Another set of differences is related to the organizational aspects of \bsd\ development when compared to commercial \os s.
The number of engineers working on similar \oss\ and closed-source projects can differ by order of magnitude.
\textsc{BSD} family \os s have hundreds of committers (see~\Cref{tab:bsd_family_overview}).
We know from grey literature that the team developing Microsoft Windows {2000} contained \num{3100} engineers responsible for developing and testing the \os~\cite{lucovsky_2000}.
For Windows Server {2003}, the number of engineers reached \num{4400}~\cite[p.~\romannum{33}]{build_master_2005}.
The team size has likely increased in the last \num{20} years.

Testament to the wide adoption of \bsd\ in the consumer space (in addition to traditionally being thought of as a server software) is the fact that Apple's closed-source \os s such as iOS, macOS, watchOS base themselves on \bsd~\cite{levin_ios_2017,singh_mac_2016}.

\subsection{Terminology}
\label{subsec:terminology}

\subsubsection{Kernel and user mode}

Various definitions for \krn\ exist in textbooks about \os s.
The \krn\ is defined as
\say{single binary program}~\cite[p.~94]{tanenbaum_operating_1997},
\say{provider of services}~\cite[p.~18]{beck_linux_1998},
\say{interface between the hardware and the software}~\cite[p.~62]{lucas_absolute_freebsd_2002},
\say{core operating system code}~\cite[p.~2]{halvorsen_os_2011},
\say{[t]he most important program}~\cite[p.~8]{bovet_understanding_2003},
\say{minimal facilities necessary for implementing additional operating-system services}~\cite[p.~22]{mckusick_design_2015},
\say{nucleus which contains the most frequently used functions in the OS, at a given time, other portions of the OS currently in use}~\cite[p.~53]{stallings_operating_2009},
or
\say{\emph{supervisor}, \emph{core}, or \emph{internals} of the operating system}~\cite[p.~4]{love_linux_2005}.

In this paper, we use the following definition: \say{[t]he kernel is the part of the system that runs in protected mode and mediates access by all user programs to the underlying hardware (e.g., CPU, keyboard, monitor, disks, network links) and software constructs (e.g., filesystem, network protocols)}~\cite[p.~22]{mckusick_design_2015}.

Code running as part of the \krn\ is called \emph{\krn\ mode}.
\say{Kernel mode refers to a mode of execution in a processor that grants access to all system memory and all CPU instructions}~\cite[p.~17]{russinovich_windows_2012}.
Sometimes \krn\ mode is also called \emph{supervisor mode}, \say{where everything is allowed} and \say{where the processor regulates direct access to hardware and unauthorized access to memory}~\cite[p.~18]{rubini_linux_2001}.
\emph{The consequences of defects in \krn\ mode are catastrophic}.
For example, \say{error in the kernel programming can block the entire system}~\cite[p.~18]{beck_linux_1998}.

The counterpart to the \krn\ mode is the \emph{user mode}.
User mode is traditionally associated with the abstraction level at which code executes.
It is not necessarily related to how the source tree layout is structured.
Therefore, we use the term \emph{\nonkrn\ code} to describe all the source that is not part of the \krn.
The definitions of user mode vary from describing it as a specific type of execution environment to a more technical scope.
For example,
\say{programs running outside the kernel}~\cite[p.~298]{tanenbaum_operating_1997},
\say{[c]ompilers and editors run in user mode}~\cite[p.~3]{tanenbaum_operating_1997}
\say{certain areas of memory are protected from the user's use and in which certain instructions may not be executed}~\cite[p.~58]{stallings_operating_2009},
\say{all pages in the user address space are accessible from user mode}~\cite[p.~17]{russinovich_windows_2012},
or a container where \say{a thread executes application code with the machine in nonprivileged protection mode}~\cite[p.~90]{mckusick_design_2015}.

User mode is sometimes referred to as \emph{user space}~\cite[p.~18]{rubini_linux_2001},
\emph{user-space}~\cite[p.~4]{love_linux_2005},
or \emph{userspace}~\cite[p.~11]{mauerer_professional_2008}.
Another way to look at user mode is \say{[i]ndividual processes exist independently alongside each other and cannot affect each other directly}~\cite[p.~18]{beck_linux_1998} or
\say{[s]oftware that runs in user space has no direct access to hardware}~\cite[p.~2]{halvorsen_os_2011}.
Limitations related to execution and resource access are the overall defining factor for user mode.
User mode applications are said to see a \say{subset of the machine's available resources and cannot perform certain system functions, directly access hardware, or otherwise misbehave}~\cite[p.~4]{love_linux_2005}.
The term \emph{userland} \say{is the nomenclature typically preferred by the BSD community for all things that do not belong to the kernel}~\cite[p.~11]{mauerer_professional_2008}.

Though some sources define an \os\ as a \say{portion of the software that runs in kernel mode or supervisor mode}~\cite[p.~3]{tanenbaum_operating_1997}, such definition is outdated.
We use a more inclusive understanding where an \os\ is defined as a \say{layer of software that manages a computer's resources for its users and their applications}~\cite[p.~6]{anderson_operating_2014}.
Modern \os s such as Microsoft Windows enable execution of critical services and device drivers in user mode in addition to \krn\ mode~\cite{orwick_developing_2007}.

\subsubsection{Committers, contributors, and maintainers}
\label{subsubsec:committers}

The \bsd\ development process and \oss, in general, have clearly defined roles for engineers contributing code.
Using definitions from \fbsd\footnote{\protect\url{https://wiki.freebsd.org/BecomingACommitter}}, the roles are as follows:

\begin{itemize}
    \item \emph{Contributor} -- an individual who is contributing code.
    \item \emph{Committer} -- an individual with \emph{write access} to the source code repository.
    \item \emph{Maintainer} -- a committer who maintains a particular subsystem.
\end{itemize}

As an alternative definition~\cite[p.~14--15]{mckusick_design_2015}, the roles are divided into
\begin{enumerate*}[label=(\alph*),before=\unskip{ }, itemjoin={{, }}, itemjoin*={{, and }}]
    \item \emph{developers} (\say{able to access the source-code repository, but they are not permitted to change it})
    \item \emph{committers} (\say{permitted to make changes to those parts of the source-code repository in which they have been authorized to work})
    \item \emph{core team} (a subset of committers who act as \say{final gatekeepers of the source code}).
\end{enumerate*}

In this paper, we do not distinguish between various roles.
From a data mining point of view, we cannot determine \emph{who exactly authored the code changes} based purely on source code history.
When a contributor submits a patch and the patch gets reviewed and approved, the committer makes the actual code change.
There is no formal and uniform way to specify the original author.
Phrases such as \say{patch from \textit{alias@domain.com},} \say{patch by \textit{alias@domain.com},} or \say{from \textit{alias}} are sometimes used.

In addition, \cor s are not mandatory for committers.
Committers are \say{required to have any nontrivial changes reviewed by at least one other person before committing them to the tree}~\cite[p.~14]{mckusick_design_2015}.
Each committer can determine the interpretation of \say{nontrivial.}
However, \cor s do not always have to be public and can be conducted by email, \textsc{IRC}, or other mechanisms.\footnote{\protect\url{https://docs.freebsd.org/en/articles/committers-guide/}}
As a result, we have only access to a subset of all \cor s.
We list this explicitly as a threat to validity in~\Cref{sec:threats}.

\section{Study Design}

\subsection{Choice of data}

\begin{table*}[ht]
  \centering
  \caption{Overview of \bsd\ family \os s. Commit hashes used to calculate \loc\ are based on {G}it{H}ub mirrors. Number of committers is as of February 10, 2022.}
  \label{tab:bsd_family_overview}
  \begin{tabular}{lrrrrrl}
    \toprule
    Project & Established & Committers & Total commits & Kernel \loc & Non-kernel \loc & Commit hash\\
    \midrule
    DragonFlyBSD& 2003 & \num{56} & \num{36283} & \num{3553712} & \num{6795448} & 0dd847d4a4fb5725\\
    FreeBSD& 1993 & \num{393} & \num{248565} & \num{7444049} & \num{10724281} & 8a7404b2aeeb6345\\
    NetBSD& 1992 & \num{277} & \num{294140} & \num{8677777} & \num{39446639} & 557728cbecd15a53\\
    OpenBSD& 1995 & \num{167} & \num{217833} & \num{5490127} & \num{14448982} & 90c94250217113fe\\
    \bottomrule
  \end{tabular}
\end{table*}

We focus on \os s that enable a clear distinction between \krn\ and \nonkrn\ source code.
Out of the widely used \os s, only a few have open-sourced their code and history of changes to source code.
Microsoft Windows uses a closed-source model.
The Windows Research Kernel~\cite{schmidt_2010} is a subset of the Windows kernel made public to researchers.
It contains only \krn\ code and is not accompanied by the history of source code changes.
Though Apple makes some parts of the \textsc{XNU} kernel and Darwin \oss\footnote{\protect\url{https://opensource.apple.com/}}, the complete source code and its history are not accessible to researchers.

Linux source code and history of changes are public.
Though Linux is \oss, it is not suitable for this investigation.
There are hundreds of Linux distributions available (everything from Android to Ubuntu)~\cite{linux_distro_suse,linux_distro_lwn}.
However, determining what constitutes \nonkrn\ code is highly dependent on a particular Linux distribution.
Contrary to popular opinion, \say{the term Linux refers to only the kernel}~\cite[p.~3]{love_linux_2005}.
It is the nucleus \say{of an OS rather than the complete OS}~\cite[p.~3]{nutt_kernel_2001}.
Linux \say{does not include all Unix applications, such as filesystem utilities, windowing systems and graphical desktops, system administrator commands, text editors, compilers, and so on}~\cite[p.~2]{bovet_understanding_2003}.
Therefore, we exclude Linux from our analysis.

Given the limited number of popular open-source \os s under active development, we use purposive sampling to select our study samples.
We choose to target the \bsd\ family of \os s for the following reasons:
\begin{enumerate*}[label=(\alph*),before=\unskip{ }, itemjoin={{, }}, itemjoin*={{, and }}]
    \item extensive and well-documented history of \os s' evolution
    \item source tree that contains a complete \os
    \item similar nature of the \os s to each other
    \item in case of \fbsd\ the availability of detailed information about \cor s over an extended period.
\end{enumerate*}

The \os s we gather data for are \dbsd, \fbsd, \nbsd, and \obsd.
\Cref{tab:bsd_family_overview} describes the characteristics of each \os, the number of commits we process, and the size of the code base.
The average age of these \os s is \num{26} years.
We calculate the \loc\ per \os\ and category of code using \texttt{scc}~\cite{boyter_scc}.
We collect the number of committers per \os\ (\dbsd\footnote{\protect\url{https://www.dragonflybsd.org/team/}},
\fbsd\footnote{\protect\url{https://docs.freebsd.org/en/articles/contributors/#staff-committers}},
\nbsd\footnote{\protect\url{http://www.netbsd.org/people/developers.html}}, and
\obsd\footnote{\protect\url{https://marc.info/?l=openbsd-announce&m=163422237101753&w=2}}) from publicly available data.

Only \fbsd\ uses a formal code collaboration tool (\phab) to conduct \cor s out of these \os s.
The rest of the \os s use a \cor\ model where not all the \cor s are public.
Therefore, the author's identity is not  always \emph{formally} tracked (see~\Cref{subsubsec:committers}).
Only the identity of the committer is recorded as an author of changes.

\subsection{Data extraction}

Source code repositories for each version of \bsd\ use several different revision control systems: \textsc{CVS}, Git, Mercurial, and Subversion.
To avoid switching between different environments and develop multiple versions of the same toolset, we utilize the fact that each \os\ has an up-to-date GitHub mirror.
To verify the validity of the mirror image, we select a random sample of \num{50} commits from each \os\ and compare the data from the GitHub mirror to the changes in the original revision control system.
We do not observe any discrepancies in the data.
We use Git commands with custom shell scripts to fetch the information about the commit history.
The collected data is parsed by an application developed in {C\#}.

\fbsd\ \cor s are performed using a code collaboration tool called \phab.
The \phab\ instance\footnote{\protect\url{https://reviews.freebsd.org/}} used to review \fbsd\ code has \cor\ data starting from 2013.
We use {P}habry~\cite{phabry_2021} to extract the information about \cor s.
The extracted data is exposed in \textsc{JSON} format.
We develop a custom application written in {C\#} to parse the \textsc{JSON} data.
All the statistical analysis is performed using custom code written in {R}.

\subsection{Selection and elimination criteria}
\label{subsec:selection_and_elimination}

\subsubsection{Filtering commits}

We focus on the main branch used for the development of each \os.
Depending on the \os, it is called \texttt{main}, \texttt{master}, or \texttt{trunk}.
We use the \texttt{{-}{-}first-parent} option for \texttt{git log} commands to have a linear commit history
We remove all the commits that do not have any code changes.
The empty commits result from either commits containing only the binary files or \say{placeholder commits} (a side-effect of converting the source code change history from systems like \textsc{CVS} to {G}it{H}ub mirror).
Our initial dataset contained \initialcommitcount\ commits, and after filtering, it decreased to \finalcommitcount\ commits.

\subsubsection{Filtering \cor s}

We perform a thorough filtering process to ensure a valid comparison between \krn\ and \nonkrn\ \cor s.
The \phab\ instance for \fbsd\ contained \initialfreebsdcrcount\ \cor s during our data collection process.
We fetch all the \cor s that are accessible to registered users.
We use only the \cor s that have gone through the entire \cor\ cycle, i.e., they were published, accepted, and eventually merged to the target branch.
The \cor s that are abandoned, ignored, or still open are not suitable for analysis because we cannot calculate metrics such as \ttm\ (see~\Cref{sec:RQ2} for definitions).
We removed the \cor s where the author \say{self-accepted} the changes, and no other engineer was involved in the \cor\ process.
To ensure that a \cor\ contains actual code changes, we eliminate the reviews where the only content is binary files.
As an additional constraint that enforces validity, we require that the \ttfr\ is before \tta, and \ttm\ is after \tta.
That restriction filters out \cor s submitted and accepted immediately without an actual review being conducted.
After applying these selection criteria, our final dataset\footnote{\protect\url{https://figshare.com/s/467523b4c41b51e80d7e}} contains \finalfreebsdcrcount\ \cor s between 2013 and 2021.

\subsection{Statistical analysis}

\subsubsection{Characteristics of data}

We report statistically significant results at a $p < .05$ and use \textsc{APA} conventions~\cite{apa}.
Our initial observation based on the histogram of total code changes per commit is that the \cc\ per commit is not distributed normally.
By \cc\ we mean the sum of the \say{added, removed or modified} lines~\cite{munson_1983}.
We define different commit types that we use in our analysis in~\Cref{sec:commit_taxonomy}.
We use the Shapiro-Wilk test~\cite{shapiro} to confirm our observation about non-normality of commit sizes for all the commit types.
The tests confirm that commit sizes for neither \krn\ ($W = 0.085, p < .05$), \nonkrn\ code ($W = 0.061, p < .05$) or mixed code ($W = .019, p < .05$) commits are normally distributed.

We make a similar observation about the non-normality of various periods describing the \cor s: \emph{\ttfr}\ ($W = 0.12, p < .05$), \emph{\tta}\ ($W = 0.17, p < .05$), and \emph{\ttm}\ ($W = 0.25, p < .05$).
See~\Cref{sec:RQ2} for how we define these periods.
The lack of normality in data leads us to use nonparametric statistical tests to analyze the relationships between different groups.

\subsubsection{Handling outliers}

As a part of data validation, we inspect the potential outlier values~\cite{iglewicz_how_1993} for the commit dates, the total size of changes (\loc\ per commit), and the duration of \cor s.
We notice only one abnormal commit in \dbsd\ that, according to the Git log, was supposedly made in 1970\footnote{\protect\url{https://www.dragonflybsd.org/mailarchive/commits/2011-05/msg00120.html}}, but the changes were committed in 2011.
We observe that commit sizes have a non-normal (right-skewed) distribution.
Because of skewed distribution, we did not apply techniques such as Tukey $1.5\times IQR$ fence exclusion criteria~\cite{tukey_1981} to attempt to remove the potential outliers~\cite{hubert_2008}.
To investigate the potential outliers in more detail,
we manually inspect a stratified sample of \num{100} commits where the size of a commit in \loc\ is more than \num{100000}.
Similarly, we analyze \num{100} \cor s where \ttm\ exceeds a month to detect abnormalities.
None of the changes were \say{inconsistent with the remainder of that set of data}~\cite[p.~4]{gladitz_barnett_1988} or \say{surprising or discrepant
to the investigator}~\cite{beckman_outlier_1983}.
We do not classify any of those changes as outliers based on our findings.
We conduct our analysis using the complete dataset that remains after applying all the filters described in~\Cref{subsec:selection_and_elimination}.

\section{RQ1: are code churn characteristics different between \krn\ and \nonkrn\ code?}
\subsection{Commit taxonomy}
\label{sec:commit_taxonomy}

For the \bsd\ family of \os s, the distinction in the \emph{location} between \krn\ and \nonkrn\ source code is clearly defined.
All the kernel source files reside under the \texttt{sys} directory (relative to the root of the source tree)~\cite{freebsd_kernel_source}.
We use that fact to categorize each commit as follows:

\begin{itemize}
	\item \textbf{Kernel}.
	Commit contains changes pertaining \emph{only} to \krn\ code.

	\item \textbf{Non-kernel}.
	Commit contains changes related \emph{only} to \nonkrn\ code.

	\item \textbf{Mixed}.
	Commit contains changes to \emph{both} \krn\ and \nonkrn\ code.
\end{itemize}

The categorization applies only to the \emph{location} of the source code.
It does not mean that a commit contains code that is \emph{executed} only at a specific level of abstraction during the runtime.
For example, a user mode application can include headers describing \krn\ data structures to pass correct parameters to a syscall.

\begin{table}[ht]
  \centering
  \caption{Commit type percentages per \os.}
  \label{tab:bsd_commit_types}
  \begin{tabular}{lrrrr}
    \toprule
    Project & Commits & Kernel & Non-kernel & Mixed\\
    \midrule
DragonFlyBSD& \num{36283}
& 50.91\%
& 44.22\%
& 4.87\%
\\
FreeBSD& \num{248565}
& 52.44\%
& 45.10\%
& 2.46\%
\\
NetBSD& \num{294140}
& 52.68\%
& 46.04\%
& 1.27\%
\\
OpenBSD& \num{217833}
& 37.43\%
& 61.11\%
& 1.46\%
\\
    \bottomrule
  \end{tabular}
\end{table}

The distribution of commits is displayed in~\Cref{tab:bsd_commit_types}.
The majority of commits contain either \krn\ or \nonkrn\ code.
Mixed commits represent, on average, only \num{2.5}\% of the population.
We can observe that this trend is consistent across different \os s.

\subsubsection{Mixed commits}
\label{subsec:mixed_commits}

To further understand the composition of mixed commits, we investigate these commits in more detail.
We use stratified random sampling to select \num{50} mixed commits for each \os\ resulting in \num{200} samples.
We then manually analyze the contents of each commit and record the presence of the following attributes: \emph{\krn\ code}, \emph{\nonkrn\ code}, \emph{documentation} (e.g., content for a \texttt{man} page), \emph{configuration} (e.g., contents of the \os\ distribution, values for default settings), and \emph{test code} (e.g., functional or unit test code).

Out of all the possible combinations ($2^5=32$), four combinations account for \num{88.5}\% of mixed commits.

\begin{itemize}
    \item The presence of both \krn\ code and \nonkrn\ code is responsible for \num{44.5}\% of mixed commits.
    \item Both \krn\ code and documentation account for \num{29}\%.
    \item Both \krn\ code and configuration account for \num{8.5}\%.
    \item The \krn\ code with \nonkrn\ code and documentation account for \num{6.5}\%.
\end{itemize}

\begin{finding}
The majority of code changes are either in \krn\ or \nonkrn\ code.
On average, only \num{2.5}\% of code changes are a mix of these categories.
In case of mixed changes, the dominating combinations are \krn\ with \nonkrn\ code and \krn\ code with updates to accompanying documentation.
\end{finding}

An alternative approach to classification is to ignore all the auxiliary changes because they are not code per se.
We choose to categorize them separately.
Based on our experience with \os\ development, a similar amount of rigor and effort goes into reviewing changes to configuration and documentation, as to reviewing source code.

\subsection{Commit size and change ratio characteristics}

\begin{table*}[ht]
  \centering
  \caption{Commit sizes per \os. N = total number, M = mean, Mdn = median, SD = standard deviation.}
  \label{tab:bsd_commit_sizes}
  \begin{tabular}{lrrrrrrrrrrrrrrr}
    \toprule
    &\multicolumn{4}{c}{Kernel}&
    \multicolumn{4}{c}{Non-kernel}&
    \multicolumn{4}{c}{Mixed}\\
    \cmidrule(lr){2-5}
    \cmidrule(lr){6-9}
    \cmidrule(lr){10-13}
    Project&\textit{N}&\textit{M}&\textit{Mdn}&\textit{SD}&\textit{N}&\textit{M}&\textit{Mdn}&\textit{SD}&\textit{N}&\textit{M}&\textit{Mdn}&\textit{SD}\\
    \midrule
DragonFlyBSD&
\num{18471}& \num{472}& \num{13}& \num{9960}
&
\num{16046}& \num{1748}& \num{10}& \num{32193}
&
\num{1766}& \num{7309}& \num{92}& \num{227835}
\\
FreeBSD&
\num{130348}& \num{168}& \num{9}& \num{3611}
&
\num{112106}& \num{298}& \num{6}& \num{11085}
&
\num{6111}& \num{1777}& \num{71}& \num{14447}
\\
NetBSD&
\num{154967}& \num{186}& \num{9}& \num{6421}
&
\num{135426}& \num{2126}& \num{7}& \num{73152}
&
\num{3747}& \num{1785}& \num{69}& \num{33372}
\\
OpenBSD&
\num{81526}& \num{176}& \num{9}& \num{7828}
&
\num{133125}& \num{644}& \num{9}& \num{26662}
&
\num{3182}& \num{1801}& \num{51}& \num{65444}
\\
    \bottomrule
  \end{tabular}
\end{table*}

Extracting the contents of a diff from a revision control system and estimating \emph{precisely} the commit size are complex tasks~\cite{hofmann_2009,nugroho_2019}.
In this paper, we use \texttt{git show} with the default diff algorithm and \texttt{diffstat} to calculate the \cc\ for each commit~\cite{diffstat}.
The \texttt{diffstat} outputs the number of inserted, deleted, and modified lines based on the contents of the diff.
The sum of these lines is \emph{commit size}.
Descriptive statistics about commit sizes for different types of commits (\krn, \nonkrn, and mixed) are displayed in~\Cref{tab:bsd_commit_sizes}.

We compare our results with the existing data about \cor\ sizes (see~\Cref{subsubsec:code_churn_size}).
Each commit does not necessarily have to result from the \cor.
For example, committers in an \oss\ project are not obligated to send out a \cor\ for each change (see~\Cref{subsubsec:committers}) and can commit trivial changes without a \cor.
Therefore, only a subset of commits is reviewed by someone other than the author.
We utilize the data about \cor\ size as a close approximation because engineers tend to submit the contents of a \cor\ as a single commit.

The median commit sizes in our dataset differ from the numbers presented in the existing studies (see~\Cref{subsubsec:code_churn_size}).
Our findings are smaller than the findings from various companies and \oss\ projects.
Multiple reasons may explain this outcome:
\begin{itemize}
    \item For \krn\ and \nonkrn\ commits, the median commit size is smaller than existing data points.
    We theorize that this is because committers can produce lots of small trivial changes.
    This behavior leads to the median values describing the commit size decreasing.
    We choose a stratified sample of \num{100} commits with the size of $\le 5$ \loc\ and inspect the contents of all these commits manually.
    Our observations confirm this theory.
    \item For mixed commits, the median size is larger.
    Based on our investigation in~\Cref{sec:commit_taxonomy}, we observe that mixed changes involve different layers of abstraction.
    For example, \krn, corresponding user mode libraries, test cases, and tools code in the same commit.
    Mixed commits quite often involve updates to documentation as well.
    In our industry experience, this is an expected result.
    Making atomic changes across different subsystems requires more work and increased code churn.
    \item Differences in calculation of commit or \cor\ sizes.
    This paper uses the accepted definition of code churn, which means \say{added, removed or modified} \loc~\cite{munson_1983}.
    Without more formal details from each study, we speculate that it is possible that different methods of counting the \loc\ accounts for the variance we see.
\end{itemize}

A Kruskal-Wallis test for stochastic dominance~\cite{kruskal_use_1952} reveals that there is a statistically significant difference between the mean ranks of commit sizes for at least one pair  of commit types $(H (\num{2}) = \num{16010.20}, p < .05)$.
After performing a post hoc pairwise Dunn test~\cite{dunn_1961,dunn_1964} with a
Bonferroni correction~\cite{bonferroni}, we observe a difference in commit sizes between
all commit types ($p < .05$).

\begin{finding}
Kernel commits are bigger than \nonkrn\ commits.
Mixed commits involving both \krn\ and \nonkrn\ code are the largest.
\end{finding}

\subsection{Commit impact on code base size}
\label{subsec:impact-on-codebase-size}

The fact that the total size of the software grows over time can be easily observed in \oss\ and commercial software development.
For Linux~\cite{oded_2006,linux_kernel_2017_report,linux_kernel_foundation_report}, and FreeBSD~\cite{izurieta_2006}, the growth has been documented and studied.

\begin{figure}[!htbp]
    \centering
    \includegraphics[width=0.5\textwidth,keepaspectratio]{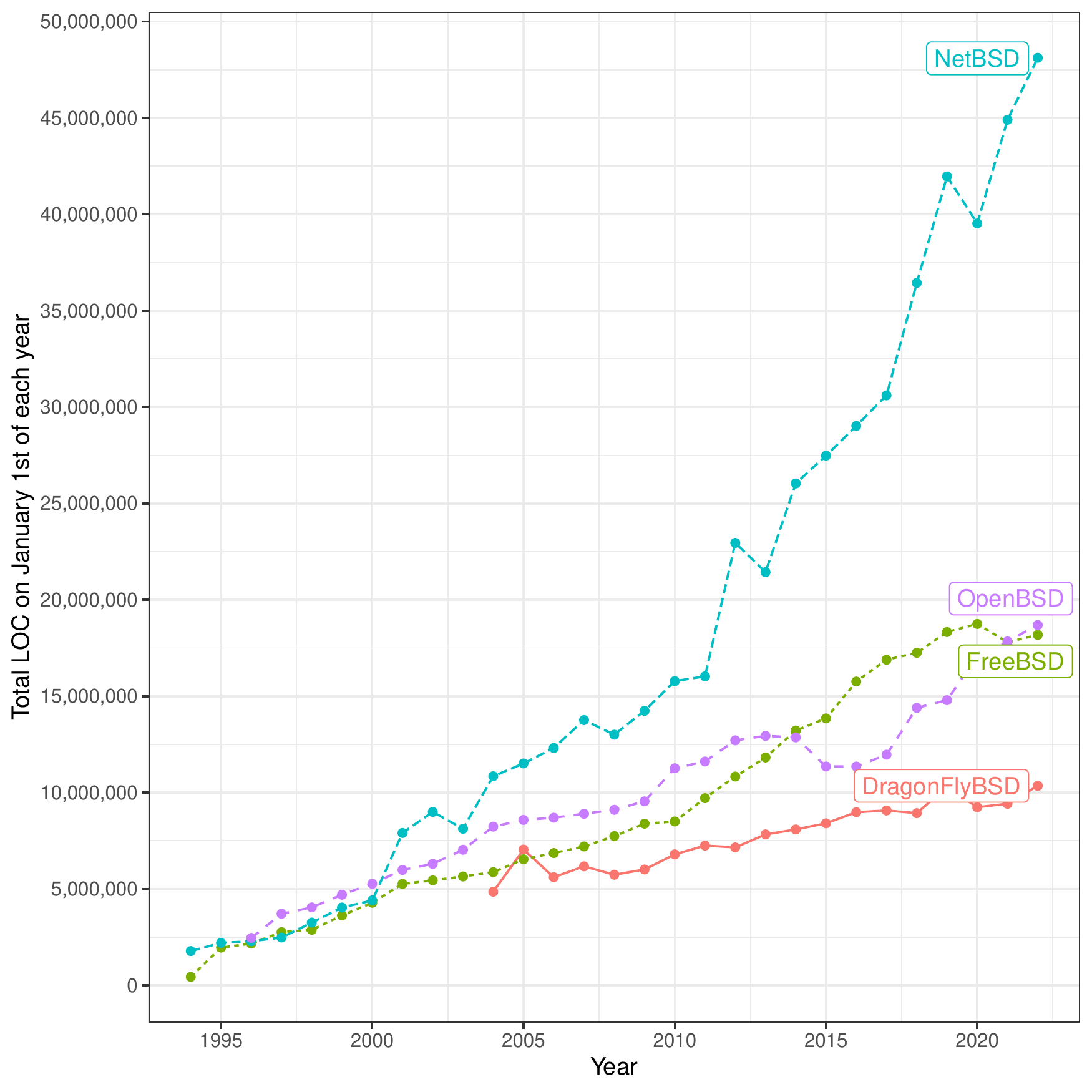}
    \caption{Evolution of the size of various \bsd\ source trees. \loc\ is calculated using the snapshot of the source tree on January 1\textsuperscript{st} of each year using \texttt{scc}.}
    \label{fig:bsd-code-growth}
\end{figure}

We can see in~\Cref{fig:bsd-code-growth} that all the \os s we study exhibit a pattern of continuous growth.
The rate of the annual increase in the code base size between \os s is similar.
The \nbsd\ growth pattern is an outlier, for which we lack an explanation.
The self-declared focus of each \os\ may partially account for the growth rates we observe.
Future studies should investigate its relationship to the growth in the size of code base.

\begin{finding}
The median annual growth rate across all \os s for \krn\ code is \num{7.15}\%.
For \nonkrn\ code, the growth rate is \num{6.19}\%.
The size of both subsystems increases at a similar rate.
\end{finding}

A limited amount of data about closed-source commercial \os s is available to the public.
Based on the grey literature, the growth trend is similar.
Microsoft Windows {\textsc{NT} 3\@.1}, released in {1993}, contained 4--5 million \loc~\cite[p.~\romannum{33}]{build_master_2005}.
In \num{10} years, the size of the code base increased \emph{by order of magnitude} to approximately \num{50} million \loc\ during the release of Windows Server {2003}~\cite[p.~\romannum{33}]{build_master_2005}.
From our industry experience, we know that the size of the Windows code base has only continued to increase since.

A reasonable hypothesis is that most commits are either adding or modifying the code, with only a tiny subset of commits reducing the size of the code base.
We do not know if that commit pattern applies equally to the \krn\ and \nonkrn\ code.
Based on our experience with commercial \os\ development, we know that each \emph{unnecessary} line in \krn\ code is generally treated as a liability due to the catastrophic consequences of \krn\ defects.
Therefore, at least amongst the \krn\ engineers, there is a firm conviction and desire to constantly reduce the size of the code base and remove any unnecessary code.

\begin{table}[ht]
  \centering
  \caption{Percentages of neutral ($=$), additive ($\uparrow$), and subtractive ($\downarrow$) commits per \os\ and commit type.}
  \label{tab:bsd_commit_neutral_incr_decr}
  \begin{tabular}{llrrr}
    \toprule
    \multicolumn{1}{l}{Project}&
    \multicolumn{1}{l}{Commit type}&
    \multicolumn{1}{c}{$=$}&
    \multicolumn{1}{c}{$\uparrow$}&
    \multicolumn{1}{c}{$\downarrow$}\\
    \midrule
DragonFlyBSD&Kernel&18.16\% & 58.22\% & 23.62\% \\
&Non-kernel&23.83\% & 53.30\% & 22.88\% \\
    \midrule
FreeBSD&Kernel&22.27\% & 55.26\% & 22.47\% \\
&Non-kernel&28.31\% & 54.48\% & 17.21\% \\
    \midrule
NetBSD&Kernel&24.99\% & 55.15\% & 19.86\% \\
&Non-kernel&29.77\% & 54.48\% & 15.75\% \\
    \midrule
OpenBSD&Kernel&22.56\% & 53.88\% & 23.56\% \\
&Non-kernel&27.44\% & 51.74\% & 20.82\% \\
    \bottomrule
  \end{tabular}
\end{table}

To verify this theory, we investigate the impact of commits on the size of the code base in more detail.
We categorize the commits into the following three categories:

\begin{itemize}
    \item \textbf{Neutral}. All code changes are modifications (i.e., changing the existing \loc) or an equal number of insertions and deletions.
    Therefore, the commit does not increase the size of the code base.
    \item \textbf{Additive}. The number of \loc\ inserted is greater than the number of \loc\ deleted.
    \item \textbf{Subtractive}. The number of \loc\ deleted is greater than the number of \loc\ inserted.
\end{itemize}

From~\Cref{tab:bsd_commit_neutral_incr_decr}, we can see that the distribution of commits is uniform across different \os s using this categorization.
We can observe
\begin{enumerate*}[label=(\alph*),before=\unskip{ }, itemjoin={{, }}, itemjoin*={{, and }}]
    \item more additive \nonkrn\ commits than \krn\ commits
    \item more subtractive \krn\ commits than \nonkrn\ commits.
\end{enumerate*}
However, the differences between the percentages of commit types for \krn\ and \nonkrn\ code are relatively minor (approximately 1--5\%).
This finding is insufficient to confirm our intuition that \krn\ commits are focused more on eliminating the dead code or reducing the size of the \krn\ code base than \nonkrn\ commits.
In theory, a small number of commits may systematically remove large amounts of \loc\ from the \krn\ code base to remove dead code.
Based on our sampling of commits that delete more than \num{1000} \loc\ in the \krn, we do not find any evidence for this.
In addition, given the general guidance for commit size in \oss\ development, we find that the frequent usage of commits containing many \loc\ to remove code is unlikely.
For example, Linux kernel guidelines state that kernel developers should \say{break their changes down into \emph{small}, logical pieces}~\cite{linux_kernel_foundation_report}.

\begin{finding}
Both \krn\ and \nonkrn\ commits have similar neutral, additive, and subtractive commit distributions.
More than 50\% of commits in both categories increase the size of the code base.
\end{finding}

\begin{table*}[ht]
  \centering
  \caption{Ratio of code changes per \os\ and commit type. N = total number of commits, M = mean, Mdn = median, SD = standard deviation.}
  \label{tab:bsd_commit_change_ratios}
  \begin{tabular}{llrrrrrrrrrr}
    \toprule
    \multicolumn{3}{c}{Project info}&
    \multicolumn{3}{c}{Insert ratio}&
    \multicolumn{3}{c}{Delete ratio}&
    \multicolumn{3}{c}{Modification ratio}\\
    \cmidrule(lr){1-3}
    \cmidrule(lr){4-6}
    \cmidrule(lr){7-9}
    \cmidrule(lr){10-12}
    Project&Commit type&\textit{N}&\textit{M}&\textit{Mdn}&\textit{SD}&\textit{M}&\textit{Mdn}&\textit{SD}&\textit{M}&\textit{Mdn}&\textit{SD}\\
    \midrule
DragonFlyBSD& Kernel &\num{18471} & 40.94\% & 40.00\% & 36.98\% & 17.99\% & 0.00\% & 29.53\% & 41.07\% & 33.33\% & 35.31\% \\
& Non-kernel &\num{16046} & 37.80\% & 28.57\% & 38.47\% & 16.34\% & 0.00\% & 29.49\% & 45.86\% & 40.00\% & 37.97\% \\
& Mixed &\num{1766} & 44.98\% & 46.86\% & 36.79\% & 22.76\% & 5.25\% & 32.56\% & 32.26\% & 21.43\% & 32.18\% \\
    \midrule
FreeBSD& Kernel &\num{130348} & 41.47\% & 36.84\% & 39.40\% & 17.73\% & 0.00\% & 30.51\% & 40.81\% & 31.71\% & 38.15\% \\
& Non-kernel &\num{112106} & 40.72\% & 33.33\% & 40.60\% & 13.17\% & 0.00\% & 27.91\% & 46.12\% & 40.00\% & 40.18\% \\
& Mixed &\num{6111} & 54.64\% & 63.93\% & 37.81\% & 16.47\% & 0.00\% & 29.15\% & 28.90\% & 16.67\% & 31.97\% \\
    \midrule
NetBSD& Kernel &\num{154967} & 33.91\% & 27.27\% & 34.06\% & 12.59\% & 0.00\% & 24.13\% & 53.50\% & 50.00\% & 34.29\% \\
& Non-kernel &\num{135426} & 34.22\% & 25.00\% & 35.44\% & 10.13\% & 0.00\% & 22.65\% & 55.65\% & 50.00\% & 35.94\% \\
& Mixed &\num{3747} & 43.86\% & 44.57\% & 36.55\% & 13.84\% & 0.00\% & 25.43\% & 42.31\% & 33.33\% & 34.88\% \\
    \midrule
OpenBSD& Kernel &\num{81526} & 35.47\% & 30.95\% & 35.01\% & 15.51\% & 0.00\% & 26.78\% & 49.02\% & 44.44\% & 34.30\% \\
& Non-kernel &\num{133125} & 33.02\% & 22.99\% & 35.17\% & 13.59\% & 0.00\% & 25.76\% & 53.40\% & 50.00\% & 35.87\% \\
& Mixed &\num{3182} & 42.60\% & 41.68\% & 36.02\% & 15.50\% & 0.00\% & 28.02\% & 41.90\% & 34.48\% & 33.49\% \\
    \bottomrule
  \end{tabular}
\end{table*}

\subsection{Ratio of code changes in a commit}

To further understand the composition of commits, we investigate the commit content in more detail.
We calculate a percentage ratio of different types of code changes (insertions, deletions, and modifications) per each commit.
For example, we define the percentage of \emph{insert ratio} as:
\[
    \frac{Total\ insertions\ (\loc)}{Commit\ size\ (\loc)}\times 100.
\]
Similarly, we calculate the \emph{delete ratio} and \emph{modify ratio}.
We present the percentages of various ratios for each \os\ and commit type in~\Cref{tab:bsd_commit_change_ratios}.
On average only 2.5\% of code changes are mixed (see~\Cref{subsec:mixed_commits}).
Our analysis is primarily interested in the differences between \krn\ and \nonkrn\ code.

As an immediate observation, we notice that the median delete ratio is 0\% for almost all the categories.
Based on the discussion in~\Cref{subsec:impact-on-codebase-size} this is an expected finding.
The only outlier behavior in \dbsd\ has a \num{5.25}\% of delete ratio in mixed commits.
To investigate this finding, we pick a random sample of \num{100} \dbsd\ commits where the delete ratio was more than \num{50}\% and analyze each commit manually.
Our analysis indicates that \num{82}\% of commits are associated with a deliberate effort to remove obsolete features\footnote{\url{https://github.com/DragonFlyBSD/DragonFlyBSD/commit/5ca0a96}} and functionality.\footnote{\url{https://github.com/DragonFlyBSD/DragonFlyBSD/commit/7c87aae}}
Therefore, mixed commits have a very high delete ratio percentage (up to \num{100}\%).
Consequently, that increases both the mean and median values for mixed commits.
The rest of the commits are associated with tasks such as refactoring, reverting previous commits, or general code hygiene\footnote{\url{https://github.com/DragonFlyBSD/DragonFlyBSD/commit/0df73a2}} related work.

We notice after observing mean and median values for the insert ratio that \krn\ commits tend to have a higher insert ratio than \nonkrn\ commits
A Mann-Whitney \textit{U} test~\cite{mann_whitney} indicated that this difference was statistically significant $U(N_{Kernel} = \num{100000}, N_{Non-kernel} = \num{100000}) = \num{10133125316.5}, z = \num{10.61}, p < .05$.
Similarly, we observe that modification ratio tends to be smaller for \krn\ code than \nonkrn\ code.
Based on Mann-Whitney \textit{U} test the difference was statistically significant $U(N_{Kernel} = \num{100000}, N_{Non-kernel} = \num{100000}) = \num{9699003058.5}, z = \num{-23.50}, p < .05$.

One possible explanation for the higher insert ratio in the \krn\ is our anecdotal observation that engineers who want to work on \os s generally tend to work on \krn\ code as opposed to \nonkrn\ code, focusing on new features and improvements to the \krn.
We find that theory plausible because it is beneficial for the \os's adoption rate.
That means code changes to support new hardware platforms, modern networking protocols, or port over existing device drivers.

For the lower modification ratio in the \krn, we speculate that engineers are more careful when changing the existing code than adding new code.
With new code, the author has detailed knowledge about why the code is being added and what it does.
The existing code, especially the \krn, which may be decades old, has a limited test coverage (if any), and there is often no context or sufficient documentation to understand the implementation details.
\emph{Making \krn\ changes under those constraints is riskier than adding new code}.
In \nonkrn\ code, the consequences of introducing defects are less severe, and therefore engineers are willing to take more risks when updating existing code.
Future studies should test that speculative intuition.

\begin{finding}
The delete ratio is abysmal across all commit types.
The insert ratio in \krn\ code is higher than in \nonkrn\ code.
The modification ratio in \krn\ code is lower than in \nonkrn\ code.
\end{finding}

\section{RQ2: is there a difference in the code velocity between \krn\ and \nonkrn\ code?}
\label{sec:RQ2}

\begin{table*}[ht]
  \centering
  \caption{Characteristics of \num{14875} \fbsd\ \cor s. M = mean, Mdn = median, SD = standard deviation. Time periods are given in hours.}
  \label{tab:bsd_code_reviews}
  \begin{tabular}{lrrrrrrrrrrrr}
    \toprule
    &\multicolumn{3}{c}{Kernel (\num{6405} reviews)}&
    \multicolumn{3}{c}{Non-kernel (\num{7573} reviews)}&
    \multicolumn{3}{c}{Mixed (\num{897} reviews)}\\
    \cmidrule(lr){2-4}
    \cmidrule(lr){5-7}
    \cmidrule(lr){8-10}
    Period&\textit{M}&\textit{Mdn}&\textit{SD}&\textit{M}&\textit{Mdn}&\textit{SD}&\textit{M}&\textit{Mdn}&\textit{SD}\\
    \midrule
Time-to-first-response & 97.86 & 4.57 & 578.13 & 107.10 & 4.16 & 682.67 & 90.17 & 4.55 & 588.79 \\
Time-to-accept & 219.18 & 11.66 & 1019.39 & 221.88 & 7.71 & 1185.22 & 445.75 & 22.43 & 1866.96 \\
Time-to-merge & 476.57 & 71.44 & 1678.20 & 547.36 & 46.03 & 2411.48 & 1013.39 & 157.41 & 2895.97 \\
    \bottomrule
  \end{tabular}
\end{table*}

We define \emph{code velocity} as the speed with which code changes are reviewed and merged into a destination branch.
Code velocity is an essential metric in the industry and is associated with engineers' job satisfaction~\cite{savor_2016,kononenko_2016}.
In environments using \textsc{CI}/\textsc{CD} (e.g., Facebook), the code velocity is essential to the entire development process~\cite{feitelson_2013}.
To investigate the code velocity in \fbsd, we use the metrics previously found to be meaningful in the industry.

\begin{itemize}
	\item We define \emph{\ttfr}\ as the time from publishing the \cor\ to the first interaction on the \cor\ by someone else than the author (excluding automated bots).
	An existing study from Microsoft that researches challenges encountered during the \cor\ process indicates that delayed \emph{response time} is the number one concern~\cite{macleod_2018}.
	Another Microsoft study about code velocity identifies two points in time that engineers consider critical: \emph{the first comment} or \emph{sign-off} from a reviewer and when the \cor\ has been marked as \emph{completed}~\cite{bird_2015}.
	\item We define \emph{\tta}\ as the time from publishing the \cor\ to when someone else than the author accepts the code changes. Once accepted, the changes are ready to be merged into the target branch.
	\item We define \emph{\ttm} as the time from publishing the \cor\ to when changes are merged to the target branch.
	A study that focuses on \cor\ performance in Xen hypervisor finds that \emph{time-to-merge} is a crucial metric to help to investigate the delays caused by the \cor\ process~\cite{izquierdo-cortazar_2017}.
\end{itemize}

We calculate these metrics for each type of commit (\krn, \nonkrn, and mixed).
We present an overview of various code review
periods in~\Cref{tab:bsd_code_reviews}.
The median size of \fbsd\ \cor\ is \num{17} \loc.
An initial observation we make is that initial engagement from reviewers is similar.
Median \ttfr\ for all commit types is  $\pm$ 1 hour.
After that, the completion of various \cor\ milestones significantly diverges between commit types.
Comparison between medians shows that while \ttfr\ for \krn\ \cor s is only \num{9.85}\% longer than for \nonkrn\ commits,
\tta\ is \num{51.23}\%, and \ttm\ is \num{55.2}\% longer.

A Kruskal-Wallis test for stochastic dominance reveals that there is a statistically significant difference between the mean ranks of \ttfr, \tta, and \ttm\ for at least one pair of commit types.
For \ttfr\ the test returned $(H (\num{2}) = \num{6.42}, p < .05)$,
for \tta\ $(H (\num{2}) = \num{165.81}, p < .05)$, and
for \ttm\ $(H (\num{2}) = \num{344.16}, p < .05)$.
After performing a post hoc pairwise Dunn test with a Bonferroni correction, we observe a difference between all commit types for \tta\ and \ttm\ ($p < .05$) and only between \krn\ and \nonkrn\ commits for \ttfr\ ($p < .05$).

\begin{finding}
Different phases of the \cor\ (\ttfr, \tta, and \ttm) take longer for the \krn\ code than for \nonkrn\ code.
Mixed \cor s take the most time to be accepted and merged.
\end{finding}

This finding feels intuitively correct.
It is reasonable to assume that reviewing \krn\ code requires greater care because of the consequences of potential defects and the technical level of knowledge required.
Mixed commits contain approximately in \num{44.5}\% cases code from both \krn\ and \nonkrn\ code.
These commits may need more reviewers (e.g., a maintainer for each affected subsystem) or require a reviewer who understands the nuances of multiple components.

Existing data about \fbsd\ code velocity is limited.
\fbsd\ study~\cite{zhu_2016} finds that the median resolve time (comparable to \ttm) is \num{23} hours and approximately \num{6} hours for the \ttfr.
The times for \ttfr\ are comparable to what we present in~\Cref{tab:bsd_code_reviews}.
The values for \ttm\ are longer for the \cor s we analyzed.
Several reasons can cause the differences.
For example, we compare our data from 2013 to 2021 to \cor s from 1995 to 2006.
Another reason is that we focus only on \cor s that were accepted and merged.
That restriction is necessary to calculate the values for \tta\ and \ttm.
That means ignoring the category of patches that the study classified as \emph{resolved} or \emph{closed}.

\section{Discussion}

\emph{Code churn characteristics}
We find that, on average, approximately 49\% of code changes are in \nonkrn\ code (see ~\Cref{tab:bsd_commit_types}).
This finding suggests that to meaningfully contribute to \os\ development, an engineer does not have to work exclusively on the \krn\ or know the intrinsic details of the development of an \os.
The fact that the insert ratio in \krn\ code changes is bigger than in \nonkrn\ code changes implies innovative opportunities (e.g., implement a new feature) to contribute to the \krn\ code base.
\emph{A surprising finding is that \krn\ code changes are bigger than \nonkrn\ code changes}.
This contradicts what we have observed during the development of commercial \os s.
The \cor s in \krn\ taking longer than \nonkrn\ \cor s is an expected finding and makes intuitive sense.

We find similar patterns related to commit sizes and taxonomy emerging across all different \os s we investigate.
Our findings about \krn\ \cor s taking longer than \nonkrn\ \cor s match with our industry experience when working on the development of commercial \os s.
Findings about mixed changes are also in agreement with our past experiences.

Though we expected all \os s to have annual code growth, \emph{the size of growth in \obsd\ during the last \num{10} years is surprising to us}.
\obsd\ has historically identified itself as an \os\ focused on leanness and security~\cite{lucas_absolute_openbsd_2003}.
These goals encourage developers to constantly \say{prune} (i.e., delete) and polish their source code to reduce the attack surface~\cite{unangst_pruning_2015}.
Nevertheless, the code base size has doubled in the last \num{10} years.

\emph{Drawing conclusions based on system-wide metrics}
When characterizing a particular company or a more extensive system, researchers need to be careful with interpreting the results and drawing conclusions from them.
For example, companies like Apple, Facebook, Google, and Microsoft have tens of complex products in development.
Each of them may evolve at a different cadence or speed.
Researchers need to be more precise when describing various evolutionary characteristics of the code base and fine-tune the scope of inquiry.

\emph{Separation of abstractions layers}.
One of the findings in our study is that mixed code changes represent a small fraction of all the changes.
Intuitively, that makes sense based on inspecting a sample of mixed commits and our industry experiences.
Code contributions spanning different components are rarer than code changes made to each component in separation.
The mixed changes also take the longest to review and contain the largest amount of code.
We speculate that longer reviews times are mainly caused because
of the size of changes,
lack of reviewers with expertise at multiple abstraction layers,
and a need to have approvals from multiple individuals.
Based on our findings, we speculate that \emph{the presence of code changes from different abstraction layers is a valuable predictor to indicate that \cor s for those changes can take longer}.

Accepted practice in the industry and \oss\ is that the main branch is always kept in a working state.
The guidance for the Linux kernel is that \say{[e]ach patch should,
when applied, yield a kernel which still builds and works properly}~\cite[p.~4]{linux_kernel_foundation_report}.
Working state means both an ability to build
the code without errors and the majority of the features working.
As an implication this means that any commit cannot break this assumption.
Based on the taxonomy of commits in~\Cref{tab:bsd_commit_types} (a tiny number of commits changing \krn\ and \nonkrn\ code simultaneously), we can infer that it is possible to work efficiently on \krn\ and \nonkrn\ code in separation.
This finding implies that various components
are isolated enough to make changes in separation.

\section{Threats to validity}
\label{sec:threats}

Like any other study, the results we present in our paper are
subject to specific categories of threats.
We enumerate the threats to construct, internal,
and external validity~\cite{shull_guide_2008}.

We thoroughly validate raw data to avoid issues with \emph{construct validity} and interpretation of theoretical constructs.
We analyze the potential outlier commits to verify that we calculate code churn correctly.
We filter out \cor s where the meaningful reviews did not occur (e.g., \say{self-accepted} changes or the only changes to binary files).
We verify that \krn\ and \nonkrn\ code locations include the code that is supposed to run at that abstraction layer.

Threats to \emph{internal validity} include impact by potential unknown factors which may influence the results.
When analyzing the \cor\ related periods (e.g., \ttfr) for \fbsd, we do not have insight into all confounding variables that can influence \cor\ times.
For example, the availability of reviewers, the time-zone for authors or reviewers, or the state of a \textsc{CI} system.
No system records that data, and the lack of data may limit internal validity.
Another threat in this category is that \cor s are not required part of the \fbsd\ development process for committers.
\say{FreeBSD committers are only required to respect each other by asking for code review before committing code to files that are actively maintained by other committers}~\cite{trong_2004}.
Our analysis bases itself on a subset of all \fbsd\ \cor s.

Concerns related to \emph{external validity} focus on applying our findings in other contexts.
We do not have access to the commit history and source code of commercial \os s such as Darwin derivatives or Microsoft Windows.
We do know from grey literature~\cite{zachary_show-stopper!_1994,maguire_writing_1993,mccarthy_dynamics_1995} and our industry experience that the characteristics of a development process for commercial and \oss\ \os s are different.
For example, motivation for product development, number of engineers involved, and development methodology.
Our findings may not be applicable in that context.
A confirmatory case study by industrial researchers is necessary to test our findings in the context of commercial \os s.

\section{Conclusions and Future Work}

We conduct a large-scale study on four \bsd\ family \os s: \dbsd, \fbsd, \nbsd, and \obsd.
Based on the literature review, we are the first to explore the differences between commit sizes, commit taxonomy, and code velocity in \krn\ and \nonkrn\ code in the context of \os s development.

Our key finding is that \emph{researchers and practitioners should view a larger software system as a collection of subsystems and sub-components, not just one entity}.
Our analysis shows that when making code changes
\begin{enumerate*}[label=(\alph*),before=\unskip{ }, itemjoin={{, }}, itemjoin*={{, and }}]
    \item developers modify either \krn\ or \nonkrn\ code, but rarely code belonging to both categories
    \item the median size of commits to \krn\ code is larger than \nonkrn\ commits
    \item both \krn\ and the \nonkrn\ code bases have a similar annual growth rate
    \item in \fbsd, the \cor s for \krn\ code take longer than \cor s for \nonkrn\ code.
\end{enumerate*}

As part of our future research, we intend to focus on the following topics:

\begin{itemize}

    \item \emph{Do developers gravitate towards \krn\ as they gain more experience with \os s development}?
It is typical for new contributors joining an \oss\ project to start contributing
by making more straightforward changes
Generally, the opportunities associated with the least amount of risk are in user mode, e.g., making changes in a command-line tool.
As engineers gain more confidence and experience, do they change their focus to areas where the stakes are higher than in user mode, e.g., device drivers?
    \item \emph{Do developers mainly contribute to their abstraction layer of choice}?
It is common in \bsd\ and Linux development to have maintainers for each area (e.g., boot sequence, tracing subsystem).
One of the interesting questions is related to the distribution between \say{specialists} (engineers who contribute only to a few narrow areas) and \say{generalists} (engineers who make changes in various components).
Based on the Linux \krn\ research, we know that most engineers (\num{62}\%) who contribute to the \krn\ have a narrow specialist profile~\cite{avelino_2017}.
We do not know if this holds in the context of an entire \os.

\end{itemize}

\bibliographystyle{IEEEtran}
\bibliography{kernel-churn}

\begin{thebibliography}{10}
\providecommand{\url}[1]{#1}
\csname url@samestyle\endcsname
\providecommand{\newblock}{\relax}
\providecommand{\bibinfo}[2]{#2}
\providecommand{\BIBentrySTDinterwordspacing}{\spaceskip=0pt\relax}
\providecommand{\BIBentryALTinterwordstretchfactor}{4}
\providecommand{\BIBentryALTinterwordspacing}{\spaceskip=\fontdimen2\font plus
\BIBentryALTinterwordstretchfactor\fontdimen3\font minus
  \fontdimen4\font\relax}
\providecommand{\BIBforeignlanguage}[2]{{%
\expandafter\ifx\csname l@#1\endcsname\relax
\typeout{** WARNING: IEEEtran.bst: No hyphenation pattern has been}%
\typeout{** loaded for the language `#1'. Using the pattern for}%
\typeout{** the default language instead.}%
\else
\language=\csname l@#1\endcsname
\fi
#2}}
\providecommand{\BIBdecl}{\relax}
\BIBdecl

\bibitem{arthur_1988}
L.~J. Arthur, \emph{Software Evolution: The Software Maintenance
  Challenge}.\hskip 1em plus 0.5em minus 0.4em\relax USA: Wiley-Interscience,
  1988.

\bibitem{tripathy_2014}
P.~Tripathy and K.~Naik, \emph{A Practitioner's Approach, Software Evolution
  and Maintenance}.\hskip 1em plus 0.5em minus 0.4em\relax USA: John Wiley \&
  Sons, Inc., 2014.

\bibitem{herraiz_2013}
\BIBentryALTinterwordspacing
I.~Herraiz, D.~Rodriguez, G.~Robles, and J.~M. Gonzalez-Barahona, ``The
  evolution of the laws of software evolution: A discussion based on a
  systematic literature review,'' \emph{ACM Comput. Surv.}, vol.~46, no.~2, dec
  2013. [Online]. Available: \url{https://doi.org/10.1145/2543581.2543595}
\BIBentrySTDinterwordspacing

\bibitem{nurolahzade_2009}
\BIBentryALTinterwordspacing
M.~Nurolahzade, S.~M. Nasehi, S.~H. Khandkar, and S.~Rawal, ``The role of patch
  review in software evolution: An analysis of the {M}ozilla {F}irefox,'' in
  \emph{Proceedings of the Joint International and Annual ERCIM Workshops on
  Principles of Software Evolution (IWPSE) and Software Evolution (Evol)
  Workshops}, ser. IWPSE-Evol '09.\hskip 1em plus 0.5em minus 0.4em\relax New
  York, NY, USA: Association for Computing Machinery, 2009, p. 9–18.
  [Online]. Available: \url{https://doi.org/10.1145/1595808.1595813}
\BIBentrySTDinterwordspacing

\bibitem{munson_1983}
J.~Munson and S.~Elbaum, ``Code churn: a measure for estimating the impact of
  code change,'' in \emph{Proceedings. International Conference on Software
  Maintenance (Cat. No. 98CB36272)}, 1998, pp. 24--31.

\bibitem{hall_software_2000}
\BIBentryALTinterwordspacing
G.~A. Hall and J.~C. Munson, ``\BIBforeignlanguage{en}{Software evolution: code
  delta and code churn},'' \emph{\BIBforeignlanguage{en}{Journal of Systems and
  Software}}, vol.~54, no.~2, pp. 111--118, Oct. 2000. [Online]. Available:
  \url{https://linkinghub.elsevier.com/retrieve/pii/S0164121200000315}
\BIBentrySTDinterwordspacing

\bibitem{spinellis_software_2021}
\BIBentryALTinterwordspacing
D.~Spinellis, P.~Louridas, and M.~Kechagia, ``\BIBforeignlanguage{en}{Software
  evolution: the lifetime of fine-grained elements},''
  \emph{\BIBforeignlanguage{en}{PeerJ Computer Science}}, vol.~7, p. e372, Feb.
  2021. [Online]. Available: \url{https://peerj.com/articles/cs-372}
\BIBentrySTDinterwordspacing

\bibitem{rigby_convergent_2013}
\BIBentryALTinterwordspacing
P.~C. Rigby and C.~Bird, ``Convergent contemporary software peer review
  practices,'' in \emph{Proceedings of the 2013 9th {Joint} {Meeting} on
  {Foundations} of {Software} {Engineering}}, ser. {ESEC}/{FSE} 2013.\hskip 1em
  plus 0.5em minus 0.4em\relax Saint Petersburg, Russia: Association for
  Computing Machinery, Aug. 2013, pp. 202--212. [Online]. Available:
  \url{https://doi.org/10.1145/2491411.2491444}
\BIBentrySTDinterwordspacing

\bibitem{sadowski_modern_2018}
\BIBentryALTinterwordspacing
C.~Sadowski, E.~S\"{o}derberg, L.~Church, M.~Sipko, and A.~Bacchelli, ``Modern
  code review: A case study at {Google},'' in \emph{Proceedings of the 40th
  {International} {Conference} on {Software} {Engineering}: {Software}
  {Engineering} in {Practice}}, ser. {ICSE}-{SEIP} '18.\hskip 1em plus 0.5em
  minus 0.4em\relax Gothenburg, Sweden: Association for Computing Machinery,
  May 2018, pp. 181--190. [Online]. Available:
  \url{https://doi.org/10.1145/3183519.3183525}
\BIBentrySTDinterwordspacing

\bibitem{rossi_continuous_2016}
\BIBentryALTinterwordspacing
C.~Rossi, E.~Shibley, S.~Su, K.~Beck, T.~Savor, and M.~Stumm, ``Continuous
  deployment of mobile software at {Facebook} (showcase),'' in
  \emph{Proceedings of the 2016 24th {ACM} {SIGSOFT} {International}
  {Symposium} on {Foundations} of {Software} {Engineering}}, ser. {FSE}
  2016.\hskip 1em plus 0.5em minus 0.4em\relax Seattle, WA, USA: Association
  for Computing Machinery, Nov. 2016, pp. 12--23. [Online]. Available:
  \url{https://doi.org/10.1145/2950290.2994157}
\BIBentrySTDinterwordspacing

\bibitem{rigby_2011}
P.~C. Rigby, ``Understanding open source software peer review: Review
  processes, parameters and statistical models, and underlying behaviours and
  mechanisms,'' Ph.D. dissertation, University of Victoria, CAN, 2011,
  aAINR80365.

\bibitem{gousios_exploratory_2014}
\BIBentryALTinterwordspacing
G.~Gousios, M.~Pinzger, and A.~v. Deursen, ``An exploratory study of the
  pull-based software development model,'' in \emph{Proceedings of the 36th
  {International} {Conference} on {Software} {Engineering}}, ser. {ICSE}
  2014.\hskip 1em plus 0.5em minus 0.4em\relax Hyderabad, India: Association
  for Computing Machinery, May 2014, pp. 345--355. [Online]. Available:
  \url{https://doi.org/10.1145/2568225.2568260}
\BIBentrySTDinterwordspacing

\bibitem{macleod_2018}
L.~MacLeod, M.~Greiler, M.-A. Storey, C.~Bird, and J.~Czerwonka, ``Code
  reviewing in the trenches: Challenges and best practices,'' \emph{IEEE
  Software}, vol.~35, no.~4, pp. 34--42, 2018.

\bibitem{bird_2015}
\BIBentryALTinterwordspacing
C.~Bird, T.~Carnahan, and M.~Greiler, ``Lessons learned from building and
  deploying a code review analytics platform,'' in \emph{2015 IEEE/ACM 12th
  Working Conference on Mining Software Repositories (MSR)}.\hskip 1em plus
  0.5em minus 0.4em\relax Los Alamitos, CA, USA: IEEE Computer Society, may
  2015, pp. 191--201. [Online]. Available:
  \url{https://doi.ieeecomputersociety.org/10.1109/MSR.2015.25}
\BIBentrySTDinterwordspacing

\bibitem{izquierdo-cortazar_2017}
\BIBentryALTinterwordspacing
D.~Izquierdo-Cortazar, N.~Sekitoleko, J.~M. Gonzalez-Barahona, and L.~Kurth,
  ``Using {Metrics} to {Track} {Code} {Review} {Performance},'' in
  \emph{Proceedings of the 21st {International} {Conference} on {Evaluation}
  and {Assessment} in {Software} {Engineering}}, ser. {EASE}'17.\hskip 1em plus
  0.5em minus 0.4em\relax Karlskrona, Sweden: Association for Computing
  Machinery, Jun. 2017, pp. 214--223. [Online]. Available:
  \url{https://doi.org/10.1145/3084226.3084247}
\BIBentrySTDinterwordspacing

\bibitem{jiang_2012}
Y.~Jiang, B.~Adams, and D.~M. German, ``Will my patch make it? and how fast?:
  Case study on the {Linux} kernel,'' in \emph{Proceedings of the 10th Working
  Conference on Mining Software Repositories}, ser. MSR '13.\hskip 1em plus
  0.5em minus 0.4em\relax IEEE Press, 2013, pp. 101--110.

\bibitem{zhu_2016}
\BIBentryALTinterwordspacing
J.~Zhu, M.~Zhou, and A.~Mockus, ``Effectiveness of code contribution: From
  patch-based to pull-request-based tools,'' in \emph{Proceedings of the 2016
  24th ACM SIGSOFT International Symposium on Foundations of Software
  Engineering}, ser. FSE 2016.\hskip 1em plus 0.5em minus 0.4em\relax New York,
  NY, USA: Association for Computing Machinery, 2016, p. 871?882. [Online].
  Available: \url{https://doi.org/10.1145/2950290.2950364}
\BIBentrySTDinterwordspacing

\bibitem{tan_2019}
\BIBentryALTinterwordspacing
X.~Tan and M.~Zhou, ``How to communicate when submitting patches: An empirical
  study of the {Linux} kernel,'' \emph{Proc. ACM Hum.-Comput. Interact.},
  vol.~3, no. CSCW, nov 2019. [Online]. Available:
  \url{https://doi.org/10.1145/3359210}
\BIBentrySTDinterwordspacing

\bibitem{phacility}
\BIBentryALTinterwordspacing
Phacility, ``Phacility - {Home},'' 2021. [Online]. Available:
  \url{https://www.phacility.com/}
\BIBentrySTDinterwordspacing

\bibitem{porter_1998}
\BIBentryALTinterwordspacing
A.~Porter, H.~Siy, A.~Mockus, and L.~Votta, ``Understanding the sources of
  variation in software inspections,'' \emph{ACM Trans. Softw. Eng. Methodol.},
  vol.~7, no.~1, p. 41–79, jan 1998. [Online]. Available:
  \url{https://doi.org/10.1145/268411.268421}
\BIBentrySTDinterwordspacing

\bibitem{rigby_2013}
\BIBentryALTinterwordspacing
P.~C. Rigby and C.~Bird, ``Convergent contemporary software peer review
  practices,'' in \emph{Proceedings of the 2013 9th Joint Meeting on
  Foundations of Software Engineering}, ser. ESEC/FSE 2013.\hskip 1em plus
  0.5em minus 0.4em\relax New York, NY, USA: Association for Computing
  Machinery, 2013, p. 202–212. [Online]. Available:
  \url{https://doi.org/10.1145/2491411.2491444}
\BIBentrySTDinterwordspacing

\bibitem{winters_2020}
T.~Winters, T.~Manshreck, and H.~Wright,
  \emph{\BIBforeignlanguage{eng}{Software Engineering at {Google}: Lessons
  Learned from Programming Over Time}}, 1st~ed.\hskip 1em plus 0.5em minus
  0.4em\relax Beijing Boston Farnham Sebastopol Tokyo: O'Reilly, 2020.

\bibitem{erdamar_measuring_2021}
\BIBentryALTinterwordspacing
B.~Erdamar, ``\BIBforeignlanguage{English}{Measuring {Code} {Review} in the
  {Linux} {Kernel}},'' Master's thesis, The Technical University of Munich,
  Munich, Mar. 2021. [Online]. Available:
  \url{https://lpc.events/event/11/contributions/905/attachments/773/1795/Erdamar_2021_Measuring-Code-Review-in-the-Linux-Kernel.pdf}
\BIBentrySTDinterwordspacing

\bibitem{salus_quarter_1994}
P.~H. Salus, \emph{A quarter century of {UNIX}}.\hskip 1em plus 0.5em minus
  0.4em\relax Reading, Mass: Addison-Wesley Pub. Co, 1994.

\bibitem{mckusick_design_2015}
M.~K. McKusick, G.~V. Neville-Neil, and R.~N.~M. Watson, \emph{The Design and
  Implementation of the {FreeBSD} Operating System}, 2nd~ed.\hskip 1em plus
  0.5em minus 0.4em\relax Upper Saddle River, NJ: Addison Wesley, 2015.

\bibitem{lucas_absolute_openbsd_2003}
M.~W. Lucas, \emph{Absolute {OpenBSD}: {Unix} for the Practical
  Paranoid}.\hskip 1em plus 0.5em minus 0.4em\relax San Francisco: No Starch
  Press, 2003.

\bibitem{lucas_absolute_freebsd_2002}
------, \emph{Absolute {BSD}: the Ultimate Guide to {FreeBSD}}.\hskip 1em plus
  0.5em minus 0.4em\relax San Francisco: No Starch Press, 2002.

\bibitem{lucovsky_2000}
\BIBentryALTinterwordspacing
M.~Lucovsky. (2000) Windows --- a software engineering odyssey. [Online].
  Available:
  \url{https://www.usenix.org/legacy/events/usenix-win2000/invitedtalks/lucovsky_html/}
\BIBentrySTDinterwordspacing

\bibitem{build_master_2005}
V.~Maraia, \emph{The Build Master: {M}icrosoft's Software Configuration
  Management Best Practices}.\hskip 1em plus 0.5em minus 0.4em\relax
  Addison-Wesley Professional, 2005.

\bibitem{levin_ios_2017}
J.~Levin, \emph{\BIBforeignlanguage{eng}{*{OS} internals. {Volume} 1: {User}
  space}}, 2nd~ed.\hskip 1em plus 0.5em minus 0.4em\relax Edison, N.J:
  Technologeeks.com, 2017.

\bibitem{singh_mac_2016}
A.~Singh, \emph{\BIBforeignlanguage{English}{Mac {OS} {X} Internals: a Systems
  Approach}}.\hskip 1em plus 0.5em minus 0.4em\relax Addison-Wesley
  Professional, 2016, {OCLC}: 1005337597.

\bibitem{tanenbaum_operating_1997}
A.~S. Tanenbaum and A.~Woodhull, \emph{Operating Systems: Design and
  Implementation}, 2nd~ed.\hskip 1em plus 0.5em minus 0.4em\relax Upper Saddle
  River, NJ: Prentice Hall, 1997.

\bibitem{beck_linux_1998}
M.~Beck, Ed., \emph{\BIBforeignlanguage{eng}{Linux Kernel Internals}},
  2nd~ed.\hskip 1em plus 0.5em minus 0.4em\relax Harlow, England ; Reading,
  Mass: Addison-Wesley, 1998.

\bibitem{halvorsen_os_2011}
O.~H. Halvorsen and D.~Clarke, \emph{\BIBforeignlanguage{eng}{{OS} {X} and
  {iOS} Kernel Programming: Master Kernel Programming for Efficiency and
  Performance}}.\hskip 1em plus 0.5em minus 0.4em\relax New York, NY: Apress,
  2011.

\bibitem{bovet_understanding_2003}
D.~P. Bovet and M.~Cesati, \emph{Understanding the {Linux} Kernel},
  2nd~ed.\hskip 1em plus 0.5em minus 0.4em\relax Beijing ; Sebastopol, Calif:
  O'Reilly, 2003.

\bibitem{stallings_operating_2009}
W.~Stallings, \emph{Operating Systems: Internals and Design Principles},
  6th~ed.\hskip 1em plus 0.5em minus 0.4em\relax Upper Saddle River, N.J:
  Pearson/Prentice Hall, 2009.

\bibitem{love_linux_2005}
R.~Love, \emph{Linux Kernel Development}, 2nd~ed.\hskip 1em plus 0.5em minus
  0.4em\relax Indianapolis, Ind: Novell Press, 2005.

\bibitem{russinovich_windows_2012}
M.~E. Russinovich, D.~A. Solomon, and A.~Ionescu, \emph{Windows Internals},
  6th~ed.\hskip 1em plus 0.5em minus 0.4em\relax Redmond, Wash: Microsoft
  Press, 2012, oCLC: ocn753301527.

\bibitem{rubini_linux_2001}
A.~Rubini and J.~Corbet, \emph{Linux Device Drivers}, 2nd~ed.\hskip 1em plus
  0.5em minus 0.4em\relax Sebastopol: O'Reilly \& Associates, 2001.

\bibitem{mauerer_professional_2008}
W.~Mauerer, \emph{Professional {Linux} Kernel Architecture}, ser. Wrox
  professional guides.\hskip 1em plus 0.5em minus 0.4em\relax Indianapolis, IN:
  Wiley Pub, 2008, oCLC: ocn227198266.

\bibitem{anderson_operating_2014}
T.~Anderson and M.~Dahlin, \emph{\BIBforeignlanguage{eng}{Operating Systems:
  Principles and Practice}}, 2nd~ed.\hskip 1em plus 0.5em minus 0.4em\relax
  s.l.: Recursive Books, 2014.

\bibitem{schmidt_2010}
\BIBentryALTinterwordspacing
A.~Schmidt, A.~Polze, and D.~Probert, ``Teaching operating systems: {Windows}
  kernel projects,'' in \emph{Proceedings of the 41st ACM Technical Symposium
  on Computer Science Education}, ser. SIGCSE '10.\hskip 1em plus 0.5em minus
  0.4em\relax New York, NY, USA: Association for Computing Machinery, 2010, pp.
  490--494. [Online]. Available: \url{https://doi.org/10.1145/1734263.1734429}
\BIBentrySTDinterwordspacing

\bibitem{linux_distro_suse}
\BIBentryALTinterwordspacing
SUSE. (2022) \BIBforeignlanguage{en-US}{What is a {Linux} {Distribution}?}
  [Online]. Available:
  \url{https://www.suse.com/suse-defines/definition/linux-distribution/}
\BIBentrySTDinterwordspacing

\bibitem{linux_distro_lwn}
\BIBentryALTinterwordspacing
{Eklektix, Inc.} (2021) {T}he {LWN}.net {L}inux {D}istribution {L}ist.
  [Online]. Available: \url{https://lwn.net/Distributions/}
\BIBentrySTDinterwordspacing

\bibitem{nutt_kernel_2001}
G.~J. Nutt, \emph{Kernel Projects for {Linux}}.\hskip 1em plus 0.5em minus
  0.4em\relax Boston: Addison Wesley Longman, 2001.

\bibitem{boyter_scc}
\BIBentryALTinterwordspacing
B.~E.~C. Boyter. (2022, Mar.) {S}loc {C}loc and {C}ode (scc). [Online].
  Available: \url{https://github.com/boyter/scc/}
\BIBentrySTDinterwordspacing

\bibitem{phabry_2021}
\BIBentryALTinterwordspacing
D.~Cotet. (2021, may) Phabry. [Online]. Available:
  \url{https://github.com/dimonco/Phabry}
\BIBentrySTDinterwordspacing

\bibitem{apa}
\BIBentryALTinterwordspacing
A.~P. {Association}. (2020) \BIBforeignlanguage{en}{Publication {Manual} of the
  {American} {Psychological} {Association}, {Seventh} {Edition} (2020)}.
  [Online]. Available:
  \url{https://apastyle.apa.org/products/publication-manual-7th-edition}
\BIBentrySTDinterwordspacing

\bibitem{shapiro}
\BIBentryALTinterwordspacing
S.~S. Shapiro and M.~B. Wilk, ``\BIBforeignlanguage{en}{An analysis of variance
  test for normality (complete samples)},''
  \emph{\BIBforeignlanguage{en}{Biometrika}}, vol.~52, no. 3-4, pp. 591--611,
  Dec. 1965. [Online]. Available:
  \url{https://academic.oup.com/biomet/article-lookup/doi/10.1093/biomet/52.3-4.591}
\BIBentrySTDinterwordspacing

\bibitem{iglewicz_how_1993}
B.~Iglewicz and D.~C. Hoaglin, \emph{How to detect and handle outliers}, ser.
  {ASQC} basic references in quality control.\hskip 1em plus 0.5em minus
  0.4em\relax Milwaukee, Wis: ASQC Quality Press, 1993, no. v. 16.

\bibitem{tukey_1981}
\BIBentryALTinterwordspacing
H.~Beyer, ``\BIBforeignlanguage{en}{Tukey, {John} {W}.: {Exploratory} {Data}
  {Analysis}. {Addison}-{Wesley} {Publishing} {Company} {Reading}, {Mass}. ?
  {Menlo} {Park}, {Cal}., {London}, {Amsterdam}, {Don} {Mills}, {Ontario},
  {Sydney} 1977, {XVI}, 688 {S}.}'' \emph{\BIBforeignlanguage{en}{Biometrical
  Journal}}, vol.~23, no.~4, pp. 413--414, 1981. [Online]. Available:
  \url{http://doi.wiley.com/10.1002/bimj.4710230408}
\BIBentrySTDinterwordspacing

\bibitem{hubert_2008}
\BIBentryALTinterwordspacing
M.~Hubert and E.~Vandervieren, ``An adjusted boxplot for skewed
  distributions,'' \emph{Computational Statistics \& Data Analysis}, vol.~52,
  no.~12, pp. 5186--5201, 2008. [Online]. Available:
  \url{https://www.sciencedirect.com/science/article/pii/S0167947307004434}
\BIBentrySTDinterwordspacing

\bibitem{gladitz_barnett_1988}
\BIBentryALTinterwordspacing
V.~Barnett and T.~Lewis, ``Outliers in statistical data,'' \emph{Biometrical
  Journal}, vol.~30, no.~7, pp. 866--867, 1984. [Online]. Available:
  \url{https://doi.org/10.1002/bimj.4710300725}
\BIBentrySTDinterwordspacing

\bibitem{beckman_outlier_1983}
\BIBentryALTinterwordspacing
R.~J. Beckman and R.~D. Cook,
  ``\BIBforeignlanguage{en}{Outlier\ldots\ldots\ldots{}s},''
  \emph{\BIBforeignlanguage{en}{Technometrics}}, vol.~25, no.~2, pp. 119--149,
  May 1983. [Online]. Available:
  \url{http://www.tandfonline.com/doi/abs/10.1080/00401706.1983.10487840}
\BIBentrySTDinterwordspacing

\bibitem{freebsd_kernel_source}
\BIBentryALTinterwordspacing
{The FreeBSD Project}. (2022) \BIBforeignlanguage{en}{The layout of /usr/src}.
  Publisher: The FreeBSD Project. [Online]. Available:
  \url{https://docs.freebsd.org/en/books/developers-handbook/introduction/#introduction-layout}
\BIBentrySTDinterwordspacing

\bibitem{hofmann_2009}
\BIBentryALTinterwordspacing
P.~Hofmann and D.~Riehle, ``Estimating commit sizes efficiently,'' in
  \emph{Open Source Ecosystems: Diverse Communities Interacting, 5th {IFIP}
  {WG} 2.13 International Conference on Open Source Systems, {OSS} 2009,
  Sk{\"{o}}vde, Sweden, June 3-6, 2009. Proceedings}, ser. {IFIP} Advances in
  Information and Communication Technology, C.~Boldyreff, K.~Crowston,
  B.~Lundell, and A.~I. Wasserman, Eds., vol. 299.\hskip 1em plus 0.5em minus
  0.4em\relax Springer, 2009, pp. 105--115. [Online]. Available:
  \url{https://doi.org/10.1007/978-3-642-02032-2_11}
\BIBentrySTDinterwordspacing

\bibitem{nugroho_2019}
\BIBentryALTinterwordspacing
Y.~S. Nugroho, H.~Hata, and K.~Matsumoto, ``How different are different diff
  algorithms in {Git}?'' \emph{Empirical Software Engineering}, vol.~25, no.~1,
  pp. 790--823, Sep. 2019. [Online]. Available:
  \url{https://doi.org/10.1007/s10664-019-09772-z}
\BIBentrySTDinterwordspacing

\bibitem{diffstat}
\BIBentryALTinterwordspacing
T.~E. Dickey. (2021, mar) diffstat manual. [Online]. Available:
  \url{https://invisible-island.net/diffstat/diffstat.html}
\BIBentrySTDinterwordspacing

\bibitem{kruskal_use_1952}
\BIBentryALTinterwordspacing
W.~H. Kruskal and W.~A. Wallis, ``\BIBforeignlanguage{en}{Use of {Ranks} in
  {One}-{Criterion} {Variance} {Analysis}},''
  \emph{\BIBforeignlanguage{en}{Journal of the American Statistical
  Association}}, vol.~47, no. 260, pp. 583--621, Dec. 1952. [Online].
  Available:
  \url{http://www.tandfonline.com/doi/abs/10.1080/01621459.1952.10483441}
\BIBentrySTDinterwordspacing

\bibitem{dunn_1961}
\BIBentryALTinterwordspacing
O.~J. Dunn, ``\BIBforeignlanguage{en}{Multiple {Comparisons} among {Means}},''
  \emph{\BIBforeignlanguage{en}{Journal of the American Statistical
  Association}}, vol.~56, no. 293, pp. 52--64, Mar. 1961. [Online]. Available:
  \url{http://www.tandfonline.com/doi/abs/10.1080/01621459.1961.10482090}
\BIBentrySTDinterwordspacing

\bibitem{dunn_1964}
\BIBentryALTinterwordspacing
------, ``Multiple comparisons using rank sums,'' \emph{Technometrics}, vol.~6,
  no.~3, pp. 241--252, 1964. [Online]. Available:
  \url{https://www.tandfonline.com/doi/abs/10.1080/00401706.1964.10490181}
\BIBentrySTDinterwordspacing

\bibitem{bonferroni}
H.~Abdi, ``The bonferonni and \v{S}id\'{a}k corrections for multiple
  comparisons,'' \emph{Encyclopedia of measurement and statistics}, vol.~3, 01
  2007.

\bibitem{oded_2006}
\BIBentryALTinterwordspacing
O.~Koren, ``A study of the {Linux} kernel evolution,'' \emph{SIGOPS Oper. Syst.
  Rev.}, vol.~40, no.~2, p. 110–112, apr 2006. [Online]. Available:
  \url{https://doi.org/10.1145/1131322.1131325}
\BIBentrySTDinterwordspacing

\bibitem{linux_kernel_2017_report}
\BIBentryALTinterwordspacing
{The Linux Foundation}. (2017) \BIBforeignlanguage{en-US}{State of {Linux}
  {Kernel} {Development} 2017}. [Online]. Available:
  \url{https://www.linuxfoundation.org/tools/state-of-linux-kernel-development-2017/}
\BIBentrySTDinterwordspacing

\bibitem{linux_kernel_foundation_report}
J.~Corbet, G.~Kroah-Hartman, and A.~McPherson. (2012, 03/2012) Linux kernel
  development: How fast it is going, who is doing it, what they are doing, and
  who is sponsoring it.

\bibitem{izurieta_2006}
\BIBentryALTinterwordspacing
C.~Izurieta and J.~Bieman, ``The evolution of {FreeBSD} and {Linux},'' in
  \emph{Proceedings of the 2006 ACM/IEEE International Symposium on Empirical
  Software Engineering}, ser. ISESE '06.\hskip 1em plus 0.5em minus 0.4em\relax
  New York, NY, USA: Association for Computing Machinery, 2006, p. 204–211.
  [Online]. Available: \url{https://doi.org/10.1145/1159733.1159765}
\BIBentrySTDinterwordspacing

\bibitem{mann_whitney}
\BIBentryALTinterwordspacing
H.~B. Mann and D.~R. Whitney, ``\BIBforeignlanguage{en}{On a {Test} of
  {Whether} one of {Two} {Random} {Variables} is {Stochastically} {Larger} than
  the {Other}},'' \emph{\BIBforeignlanguage{en}{The Annals of Mathematical
  Statistics}}, vol.~18, no.~1, pp. 50--60, Mar. 1947. [Online]. Available:
  \url{http://projecteuclid.org/euclid.aoms/1177730491}
\BIBentrySTDinterwordspacing

\bibitem{savor_2016}
T.~Savor, M.~Douglas, M.~Gentili, L.~Williams, K.~Beck, and M.~Stumm,
  ``Continuous {Deployment} at {Facebook} and {OANDA},'' in \emph{2016
  {IEEE}/{ACM} 38th {International} {Conference} on {Software} {Engineering}
  {Companion} ({ICSE}-{C})}, May 2016, pp. 21--30.

\bibitem{kononenko_2016}
\BIBentryALTinterwordspacing
O.~Kononenko, O.~Baysal, and M.~W. Godfrey, ``Code review quality: How
  developers see it,'' in \emph{Proceedings of the 38th {International}
  {Conference} on {Software} {Engineering}}, ser. {ICSE} '16.\hskip 1em plus
  0.5em minus 0.4em\relax Austin, Texas: Association for Computing Machinery,
  May 2016, pp. 1028--1038. [Online]. Available:
  \url{https://doi.org/10.1145/2884781.2884840}
\BIBentrySTDinterwordspacing

\bibitem{feitelson_2013}
D.~G. {Feitelson}, E.~{Frachtenberg}, and K.~L. {Beck}, ``Development and
  deployment at {Facebook},'' \emph{IEEE Internet Computing}, vol.~17, no.~4,
  pp. 8--17, 2013.

\bibitem{unangst_pruning_2015}
\BIBentryALTinterwordspacing
T.~Unangst, ``Pruning and {Polishing}: {Keeping} {OpenBSD} {Modern},'' in
  \emph{Proceedings of {AsiaBSDCon} 2015}.\hskip 1em plus 0.5em minus
  0.4em\relax Tokyo, Japan: Tokyo University of Science, Mar. 2015. [Online].
  Available: \url{https://www.openbsd.org/papers/pruning.html}
\BIBentrySTDinterwordspacing

\bibitem{shull_guide_2008}
F.~Shull, J.~Singer, and D.~I.~K. Sjøberg,
  \emph{\BIBforeignlanguage{eng}{Guide to Advanced Empirical Software
  Engineering}}.\hskip 1em plus 0.5em minus 0.4em\relax London: Springer, 2008.

\bibitem{trong_2004}
T.~Dinh-Trong and J.~Bieman, ``Open source software development: a case study
  of {FreeBSD},'' in \emph{10th International Symposium on Software Metrics,
  2004. Proceedings.}, 2004, pp. 96--105.

\bibitem{zachary_show-stopper!_1994}
G.~P. Zachary, \emph{Show-stopper! The breakneck race to create {Windows} {NT}
  and the next generation at {Microsoft}}.\hskip 1em plus 0.5em minus
  0.4em\relax New York : Toronto : New York: Free Press ; Maxwell Macmillan
  Canada ; Maxwell Macmillan International, 1994.

\bibitem{maguire_writing_1993}
S.~Maguire, \emph{Writing solid code: {Microsoft}'s techniques for developing
  bug-free {C} programs}.\hskip 1em plus 0.5em minus 0.4em\relax Redmond, Wash:
  Microsoft Press, 1993.

\bibitem{mccarthy_dynamics_1995}
J.~McCarthy, \emph{Dynamics of software development}.\hskip 1em plus 0.5em
  minus 0.4em\relax Redmond, Wash: Microsoft Press, 1995.

\bibitem{avelino_2017}
\BIBentryALTinterwordspacing
G.~Avelino, L.~Passos, A.~Hora, and M.~T. Valente, ``Assessing code authorship:
  The case of the {Linux} kernel,'' in \emph{Open Source Systems: Towards
  Robust Practices}.\hskip 1em plus 0.5em minus 0.4em\relax Springer
  International Publishing, 2017, pp. 151--163. [Online]. Available:
  \url{https://doi.org/10.1007/978-3-319-57735-7_15}
\BIBentrySTDinterwordspacing

\end{thebibliography}

\end{document}